\documentclass[aps, prb, twocolumn, floatfix, superscriptaddress, nofootinbib]{revtex4-2}
\usepackage{amsmath, amssymb, graphicx, epsfig, xcolor}
\usepackage[colorlinks=true, allcolors=blue]{hyperref}

\begin{document}

\title {Prediction of Van Hove singularity systems in ternary borides}

\author{Yang Sun$^{\dag\ast}$}
\address{Department of Physics, Xiamen University, Xiamen 361005, China}
\address{Department of Physics, Iowa State University, Ames, IA 50011, USA}

\author{Zhen Zhang$^{\dag}$}
\address{Department of Physics, Iowa State University, Ames, IA 50011, USA}

\author{Andrew P Porter$^{\dag}$}
\address{Department of Chemistry, Iowa State University, Ames, IA 50011, USA}
\address{Ames National Laboratory, U.S. Department of Energy, Ames, IA 50011, USA}

\author{Kirill Kovnir$^{\ast}$}
\address{Department of Chemistry, Iowa State University, Ames, IA 50011, USA}
\address{Ames National Laboratory, U.S. Department of Energy, Ames, IA 50011, USA}

\author{Kai-Ming Ho}
\address{Department of Physics, Iowa State University, Ames, IA 50011, USA}

\author{Vladimir Antropov$^{\ast}$}
\address{Department of Physics, Iowa State University, Ames, IA 50011, USA}
\address{Ames National Laboratory, U.S. Department of Energy, Ames, IA 50011, USA}

\def\thefootnote{$\dag$}\footnotetext{Equal contribution.}\def\thefootnote{\arabic{footnote}}
\def\thefootnote{$\ast$}\footnotetext{Email: yangsun@xmu.edu.cn (Y.S.); kovnir@iastate.edu (K.K.); antropov@iastate.edu (V.A.)}\def\thefootnote{\arabic{footnote}}

\date{Sep. 10, 2023}

\begin{abstract}
A computational search for stable structures among both $\alpha$ and $\beta$ phases of ternary ATB$_4$ borides (A = Mg, Ca, Sr, Ba, Al, Ga, and Zn, T is $3d$ or $4d$ transition elements) has been performed. We found that $\alpha$-ATB$_4$ compounds with A = Mg, Ca, Al, and T = V, Cr, Mn, Fe, Ni, and Co form a family of structurally stable or almost stable materials. These systems are metallic in non-magnetic states and characterized by the formation of the localized molecular-like state of $3d$ transition metal atom dimers, which leads to the appearance of numerous Van Hove singularities (VHS) in the electronic spectrum. The closeness of these VHS to the Fermi level can be easily tuned by electron doping. For the atoms in the middle of the $3d$ row (Cr, Mn, and Fe), these VHS led to magnetic instabilities and new magnetic ground states with a weakly metallic or semiconducting nature. The magnetic ground states in these systems appear as an analog of the spin glass state. Experimental attempts to produce MgFeB$_4$ and associated challenges are discussed, and promising directions for further synthetic studies are formulated.
\end{abstract}

\maketitle

\section{Introduction}
The electronic density of states (DOS), in the vicinity of the Fermi level $E_f$ ($N(E_f)$), is crucial for the understanding of many properties of metallic systems. A significant $N(E_f)$ value typically leads to a broad spectrum of unusual and exciting electronic, magnetic, and structural properties. However, such large values of $N(E_f)$ simultaneously destroy the stability of the material, creating difficulties in their experimental synthesis. In some cases, losing initial stability is not a destructive factor, as it can transform the system into a stable magnetic, superconducting, or, for instance, charge density state \cite{1,2,3,4}.

A physical reason for developing large $N(E_f)$ was discussed in 1953 when Van Hove demonstrated the crucial role of topology in the electronic or phonon band structure \cite{5}. He has shown that peaks of the DOS are determined by so-called Van Hove critical points or singularities (VHS), i.e., places in the Brillouin zone (BZ), $\bf{k}$, where (for 2D systems) with energy dispersion $\varepsilon(\bf{k})$, an ordinary VHS with logarithmically diverging DOS occurs at a saddle point $\bf{k}$, determined by $\nabla_{\bf{k}}=0$. Thus, the energy surface area and the energy-band dispersion are closely related to peaks in the DOS. These VHSs are expected to play an important role in any properties of metallic materials where electrons at the Fermi level are involved.

The significance of such VHS becomes stronger in systems with lower dimensions \cite{6,7,8,9,10}. For instance, for 2D materials, the needed VHS is in the middle of the band, where the number of carriers is large. Such VHS 2D materials have been a hot topic in many areas of solid-state physics, especially after discovering high-temperature superconductivity in cuprates  \cite{11}. VHS scenario was proposed for the different types of superconductivity, phase separation, magnetic and charge instabilities, and their coexistence  \cite{12,13,14}.  Strong effects of VHS have been reported in many other systems. For instance, the anomalies of an anisotropic thermal expansion near points of electronic topological transition (induced by the corresponding VHS) have been discussed in Ref. \cite{15} In general, the effect of the proximity of the Fermi level to VHS on the kinetic and lattice properties of metals and alloys was studied in Ref. \cite{16}. The stress-driven Lifshitz transition was found in Sr$_2$RuO$_4$, where the uniaxial pressure lowered the saddle-point singularity below the $E_f$, which caused enhancement of the superconducting critical temperature  \cite{17,18,19}. Such measurement strongly implies that fermiology plays an ultimate role in mechanisms of many different orderings. For chemical applications, for instance, surface VHS can also serve as a capacitor for the electrons to enhance the contribution of the systems to O$_2$ absorption through electron transfer \cite{20}.

Below, we will focus on materials crystallizing in the YCrB$_4$ (or below 114) structural type. Such type of metal borides was first discovered in a series of RE-T-B$_4$ systems (RE=rare earth, T=transition metals) in the 1970s \cite{21,22,23}. The 114 families attracted much interest in studying $f$-electron magnetism \cite{24,25}. In the 2000s, the transition metal site in the RE-T-B$_4$ was populated by Al atoms to form RE-Al-B$_4$ with heavier and smaller RE elements (RE=Tm, Yb, Lu) \cite{26,27,28,29}. YbAlB$_4$ stimulated great research interests as the first Yb-based heavy fermion superconductor with quantum criticality \cite{30,31}. It is also predicted to be an ultra-high-temperature ceramic with outstanding thermal and structural properties  \cite{32}. While previous studies of 114 systems mainly focused on the properties caused by the RE, the structure is not limited to the RE-based compounds. If the element at A site can form the network to maintain the T dimer and match the (5,7)-membered boron rings, one would expect other stable phases in the 114 structures. Moreover, relatively isolated T dimers may create certain localized states. Such dimers can also be close to the magnetic threshold. However, we are unfamiliar with the observed magnetism in experimentally known systems with $3d$ dimers, and previous electronic structure studies ignored magnetism in similar borides  \cite{33,34}. Searching for such new stable systems can be a heavy burden for direct experimental synthesis. Recently, computational screening using electronic structure calculations analysis demonstrated its effectiveness in discovering new materials  \cite{35,36,37,38}. Below, we perform such computational search for the stable systems in 114 structural families, including the possibility of magnetism.

The paper is organized as follows. After reviewing the structures of these systems, we perform computational screening and identify $\alpha$-ATB$_4$ systems with A=Mg, Ca, Al, and T=V, Cr, Mn, Fe, Ni, and Co as a possible structurally stable family with VHS of different strengths (below, we call these VHS systems). Then, we show how the change of type of $3d$ atoms can be used to tune the strength of VHS, which generally can lead to instability of the paramagnetic Fermi-liquid state of density functional and the formation of a new quantum state. This can be a new electronic, magnetic, or structural state.  We focus on analyzing only magnetic instabilities and demonstrate the richness of possible magnetic ground states of these systems. Furthermore, we discuss our comprehensive experimental synthetic studies of these materials. While it did not yield the desired MgFeB$_4$ phase it allowed us to formulate directions for further synthetic endeavors for ATB$_4$ systems.\\

\section{Results and Discussion}
\subsection{114 structures}
\begin{figure}[t]
	\includegraphics[width=0.95\linewidth]{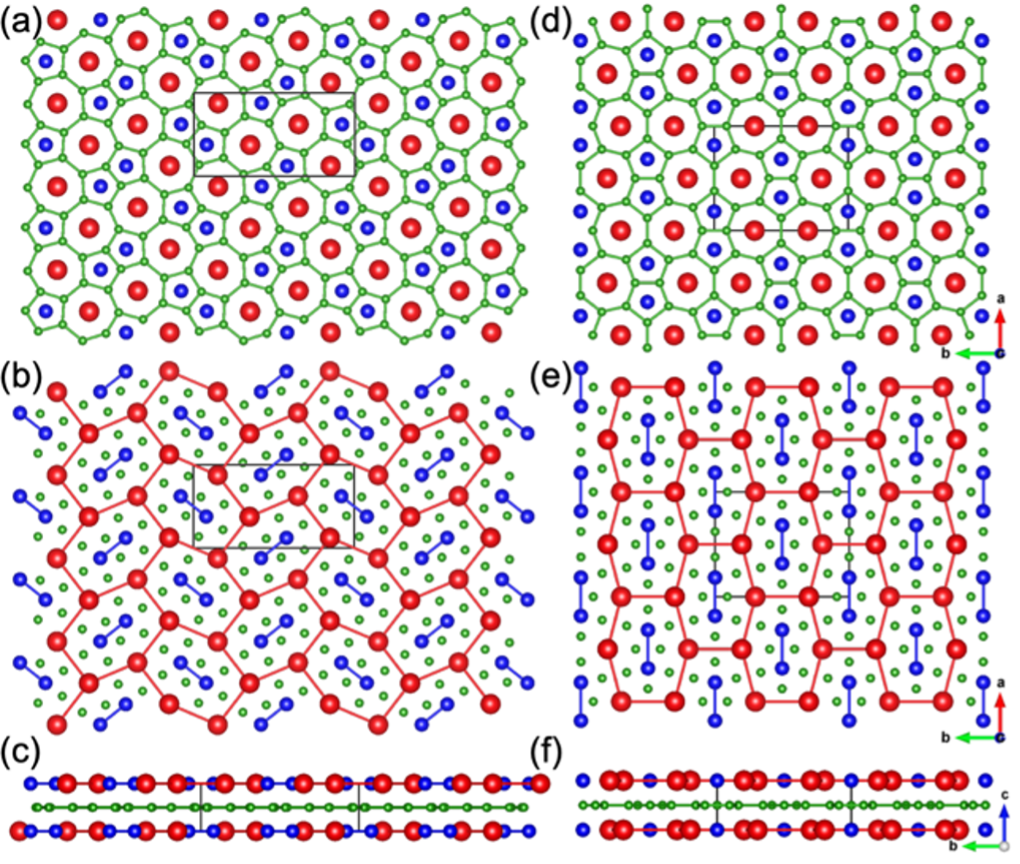}
	\caption{\textbf{Crystal structure of (a-c) $\alpha$-ATB$_4$ and (d-f) $\beta$-ATB$_4$.} (a, d) The network in the boron layer forms (5,7) rings. (b, e) The network in the metal layer. Metal A forms 6 membered-ring framework, and T forms a dimer. (c, f) Side view of metal and boron layers. Red shows the A site; Blue shows the T site; Green shows B atoms.}
	\label{fig1}
\end{figure}

114 systems have two polymorphs, i.e., $\alpha$-phase with the space group $Pbam$ and $\beta$-phase with the space group $Cmmm$. RE-T-B$_4$ mainly adopts $\alpha$-phase while RE-Al-B$_4$ can have both $\alpha$ and $\beta$ phases \cite{39}. The structure and properties of $\alpha$ and $\beta$ phases are usually similar  \cite{32}. Figure~\ref{fig1} shows their atomic packing. Both phases consist of a boron layer and a metal layer. The boron layer comprises a combination of (5,7)-membered rings. The metal layer shows a hexagonal framework of larger A atoms. Two smaller T atoms form a dimer in the center of the A framework. Along the out-plane direction, the A site corresponds to the center of the 7-membered ring in the boron layer, while the T-site aligns to the center of the 5-membered ring. Intradimer one in the $a$-$b$ plane is the closest distance between the T site atoms ($\sim$2.3-2.6 $\rm{\AA}$). The second nearest distance between the interlayer dimers, along the $c$ direction, is $\sim$40\% longer than the intradimer distance. The third nearest distance between the dimers is in the plane ($>$6 $\rm{\AA}$). Therefore, the intradimer interaction between T atoms is expected to be strongest, while interaction along the $c$-direction should be somewhat weaker. Whether or not T atoms can interact in the plane is to be found. The main difference between $\alpha$ phase and $\beta$ phase is in the orientation of in-plane networks within the $3d$ atoms layer.

\subsection{Phase stability}
\begin{figure}[t]
	\includegraphics[width=1\linewidth]{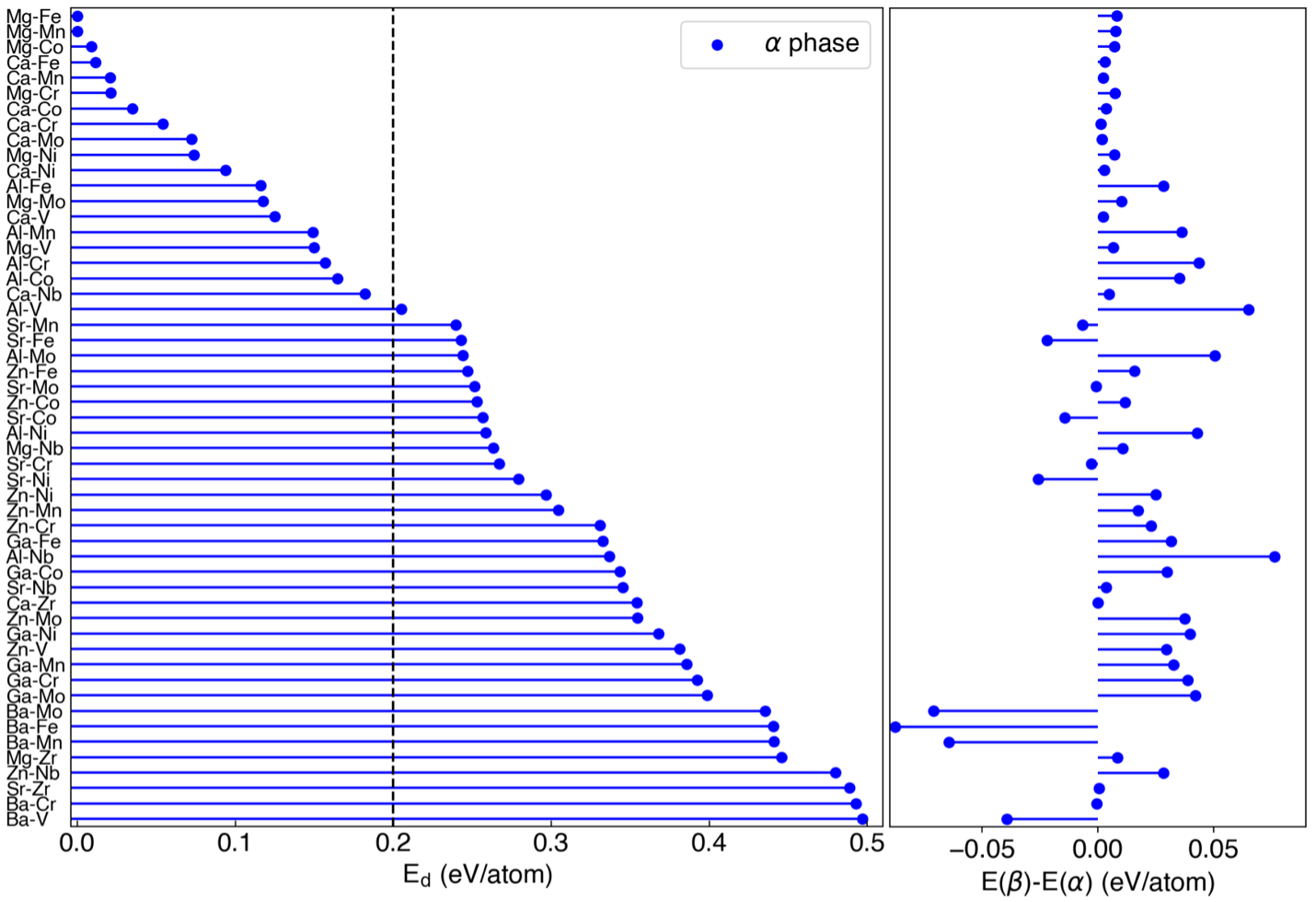}
	\caption{\textbf{Stability of non-RE ATB$_4$ phases (spin-polarized calculations).} The left panel shows the energy above the convex hull ($E_d$) for the $\alpha$ phase. The right panel shows the energy difference between $\beta$ and $\alpha$ phases. The vertical axis labels indicate specific A-T element combinations. The dashed line indicates the 0.2 eV/atom threshold to identify the metastable phases.}
	\label{fig2}
\end{figure}

The computational screening of stable phases was performed on both $\alpha$ and $\beta$ phases of ATB$_4$. We consider typical $3d$ and $4d$ transition metal elements for the T site, including V, Cr, Mn, Fe, Co, Ni, Zr, Nb, and Mo. For the A site, we consider Mg, Ca, Sr, Ba, Al, Ga, and Zn, which have relatively large ionic radii. Figure~\ref{fig2} shows the energy ($E_d$) of these ATB$_4$ phases above the convex hull formed by the compounds in existing A-T-B phase diagrams from the Material Project database  \cite{40}. Compounds with $3d$ transition metal elements are generally more stable than $4d$ ones at the T site. This can be related to the size effects that $3d$ elements fit the 5-membered boron rings more efficiently. We identified two stable phases as new ground states, MgMnB$_4$ and MgFeB$_4$. Phonon calculations confirm that MgMnB$_4$ and MgFeB$_4$ are dynamically stable (see Supplementary Fig. S1). In addition to the ground states, many low-energy metastable phases can be identified from Fig. 2. Phonon calculations also show the metastable CaMnB$_4$ and MgCoB$_4$ are dynamically stable (see Supplementary Fig. S1). If using a criterion of $E_d\sim$ 0.2 eV/atom \cite{41} to classify these metastable phases, one can see the stable and metastable phases almost only consist of $3d$ transition metals at T sites, with the A site only occupied by Mg, Ca, or Al. The $\alpha$ phases always show lower energy for these compounds than the $\beta$ phases, while their energy differences are relatively small (Fig.~\ref{fig2}).  In addition to the two stable phases, the experiments might achieve these metastable phases with  $E_d<$ 0.2 eV/atom. For instance, the LiNiB compound with $E_d=$ 0.21 eV/atom was recently synthesized from high-temperature reactions using the hydride synthetic method  \cite{42,43}.

\subsection{Electronic structures of nonmagnetic states}
\begin{figure}[b]
	\includegraphics[width=1\linewidth]{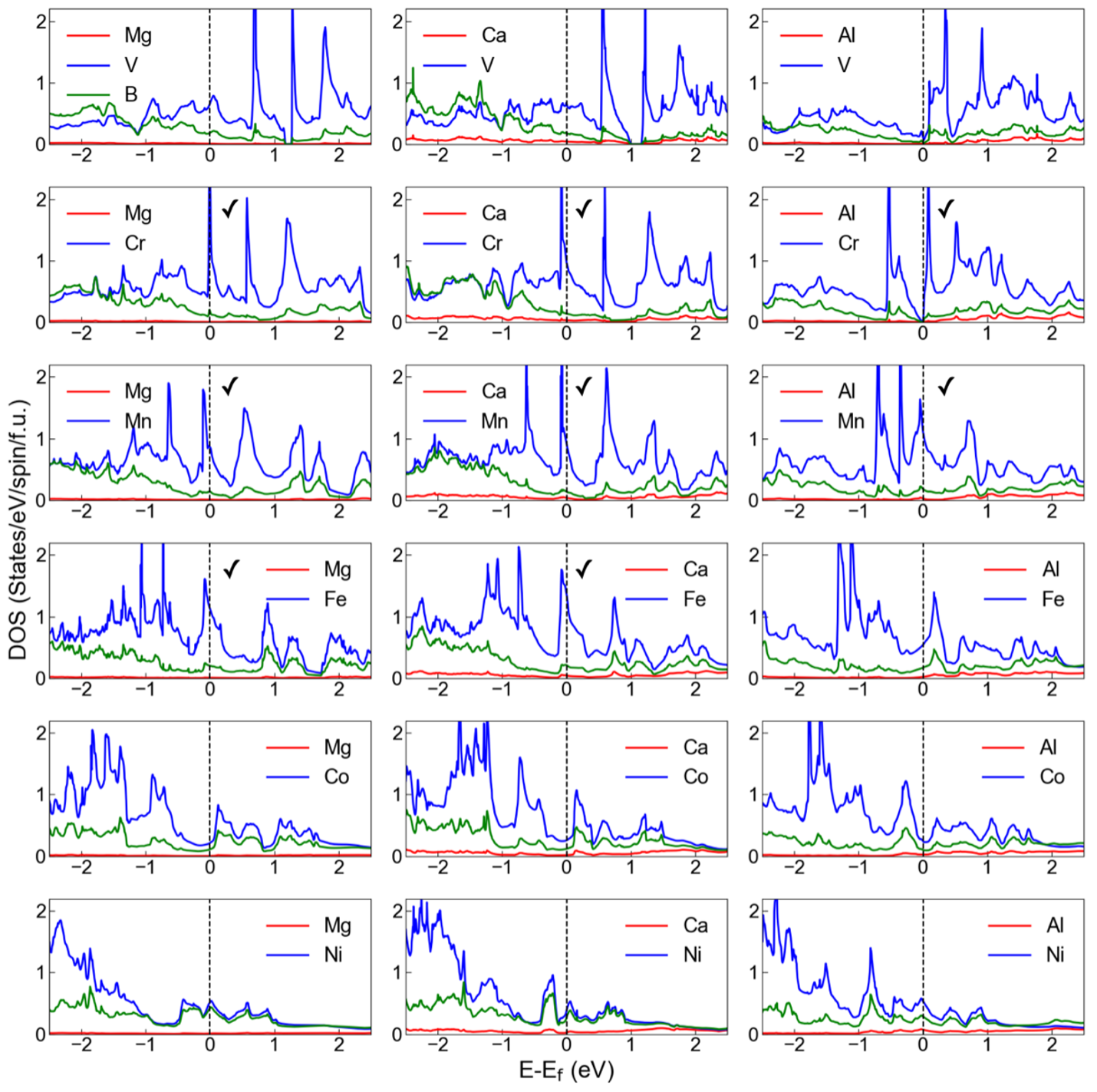}
	\caption{\textbf{The non-magnetic density of states for $\alpha$-ATB$_4$ (A=Mg, Ca, and Al; T=V, Cr, Mn, Fe, Co, Ni).} Each column has the same A element, while each row has the same T element. Contributions of states from A (red curve), T (blue curve), and B (green curve) are shown. The tick indicates the systems with a magnetic ground state.}
	\label{fig3}
\end{figure}

We now focus on stable or metastable $\alpha$-ATB$_4$ systems: A is Mg, Ca, or Al, and T is V, Cr, Mn, Fe, Ni, or Co. Their non-magnetic (NM) electronic densities of states (DOS) are shown in Fig. ~\ref{fig3}. A significant amount of VHS near the Fermi level can be identified for many systems, especially for the atoms of the middle of the $3d$-band (Cr, Mn, Fe), where the number of carriers is large. By comparing the DOS among different compounds, we find a few recurring features: First, the states near the Fermi level mainly belong to transition metal atoms forming strongly bonding dimers, with a minor contribution from B atoms and almost no contribution from Mg, Ca, or Al atoms. Second, the DOS for the different elements follows ``rigid band'' behavior. This is illustrated by Supplementary Fig. S2, which shows the integrated partial DOS for Mn in MgMnB$_4$. When we shift the Fermi level up (or down), imitating the addition (Fe) or removing (Cr) 1 electron, the resulting PDOS is very similar to the actual calculational PDOS for MgFeB$_4$ or MgCrB$_4$, respectively. From Fig.~\ref{fig3}, one can also see that when moving from V to Cr systems and further to Mn and Fe, the Fermi level is situated at or near some VHSs. These VHSs are very strong for Cr and Mn atoms, are somewhat weaker for Fe, and are inefficient for Co, Ni, and V-based 114 systems.

When the A site is occupied by Al (right column in Fig.~\ref{fig3}), the localized states of transition metals remain; however, relative positions of Fermi levels to the localized peaks are different from the Mg-based or Ca-based compounds in the same row. It is related to the various electronic populations for these systems as Al has one additional valence electron relative to the Mg or Ca system. Thus, the DOSs for the discussed family of compounds contain numerous localized dimer states above and below the Fermi level and follow rigid band behavior under doping. This observation can be verified experimentally (pending successful synthesis) as such well-defined localized states of $3d$ atomic dimers can be seen directly by spectroscopic experiments (optics in particular).

\begin{figure}[b]
	\includegraphics[width=1\linewidth]{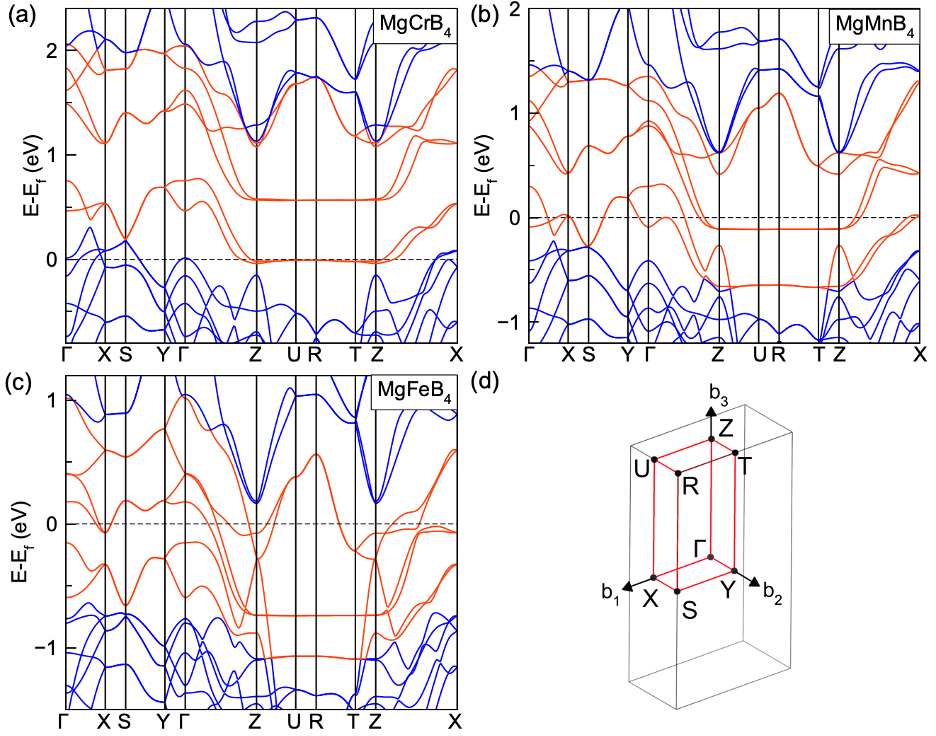}
	\caption{\textbf{Band structure for non-magnetic MgCrB$_4$, MgMnB$_4$, and MgFeB$_4$.} The inset in the left panel shows the bulk Brillouin zone. Red highlights the bands with localized states.}
	\label{fig4}
\end{figure}

VHS corresponds to the regions of the BZ where flat bands are located. In Fig.~\ref{fig4}, we show the location of such flat bands for nonmagnetic Mg-based systems. The dispersionless electronic structure at the Z-U-R-T path is persistent in all three compounds. Note that the Z-U-R-T path is parallel to the layer plane. These flat bands are the manifestation of localized electrons in the planes. According to the projected orbitals on the band structures in Supplementary Fig. S3, these flat bands have dominantly $3d$ orbital characters of transition metals with a minor contribution from B's $p$ orbital. These localized states correspond to forming quasimolecular isolated dimer states of $3d$ atoms.

The position of VHS in these systems is only sometimes precisely at the Fermi level in nonmagnetic calculations. For instance, in AlCrB$_4$, the Fermi level is practically in the gap, while the low $N(E_f)$ can also be seen for Al(V, Fe, Co)B$_4$ and (Mg, Ca)CoB$_4$, though the VHS is close to the Fermi level in all these cases. If the rigid band scenario is valid, one can predict that under suitable hole (electron) doping, one can situate the VHS at the Fermi level and study how the corresponding electronic fluctuations would change the ground state. Since the density functional theory allows us to study magnetic instabilities in these cases, we now switch to analyzing such instabilities induced by the VHS in magnetic states.

\subsection{Magnetic states}
The ferromagnetic instability is defined by fulfillment of the Stoner criteria, which indicates that for systems with $3d$ atoms, if the value of $N_{\rm{T}}(E_f)$ is around 1 eV$^{-1}$, the system is close to such instability. Figure~\ref{fig3} and Table~\ref{tab1} demonstrate that such instability exists in several of our systems. Of course, VHS in DOS can only provide information about FM instability. We must analyze corresponding Fermi-level singularities of spin susceptibility for the arbitrary magnetic instability. Below, we avoid this step by performing a direct self-consistent search of different magnetic collinear states.

\begin{table}[b]
	\caption{\textbf{The density of states at the Fermi level of transition metal (T), $N_{\rm{T}}(E_f)$ (in states/eV/spin/T), from non-magnetic calculations.}}
	\begin{ruledtabular}
		\begin{tabular}{llllll}
			Compounds  & $N_{\rm{T}}(E_f)$ & Compounds & $N_{\rm{T}}(E_f)$ & Compounds & $N_{\rm{T}}(E_f)$\\
			MgVB$_4$ & 0.667 & CaVB$_4$ & 0.674 & AlVB$_4$ & 0.162\\
			MgCrB$_4$ & 2.629 & CaCrB$_4$ & 0.979 & AlCrB$_4$ & 0.028\\
			MgMnB$_4$ & 0.865 & CaMnB$_4$ & 0.945 & AlMnB$_4$ & 1.067\\
			MgFeB$_4$ & 1.172 & CaFeB$_4$ & 1.227 & AlFeB$_4$ & 0.592\\
			MgCoB$_4$ & 0.243 & CaCoB$_4$ & 0.302 & AlCoB$_4$ & 0.251\\
			MgNiB$_4$ & 0.512 & CaNiB$_4$ & 0.351 & AlNiB$_4$ & 0.535\\
		\end{tabular}
	\end{ruledtabular}
	\label{tab1}
\end{table}

\begin{figure}[t]
	\includegraphics[width=1\linewidth]{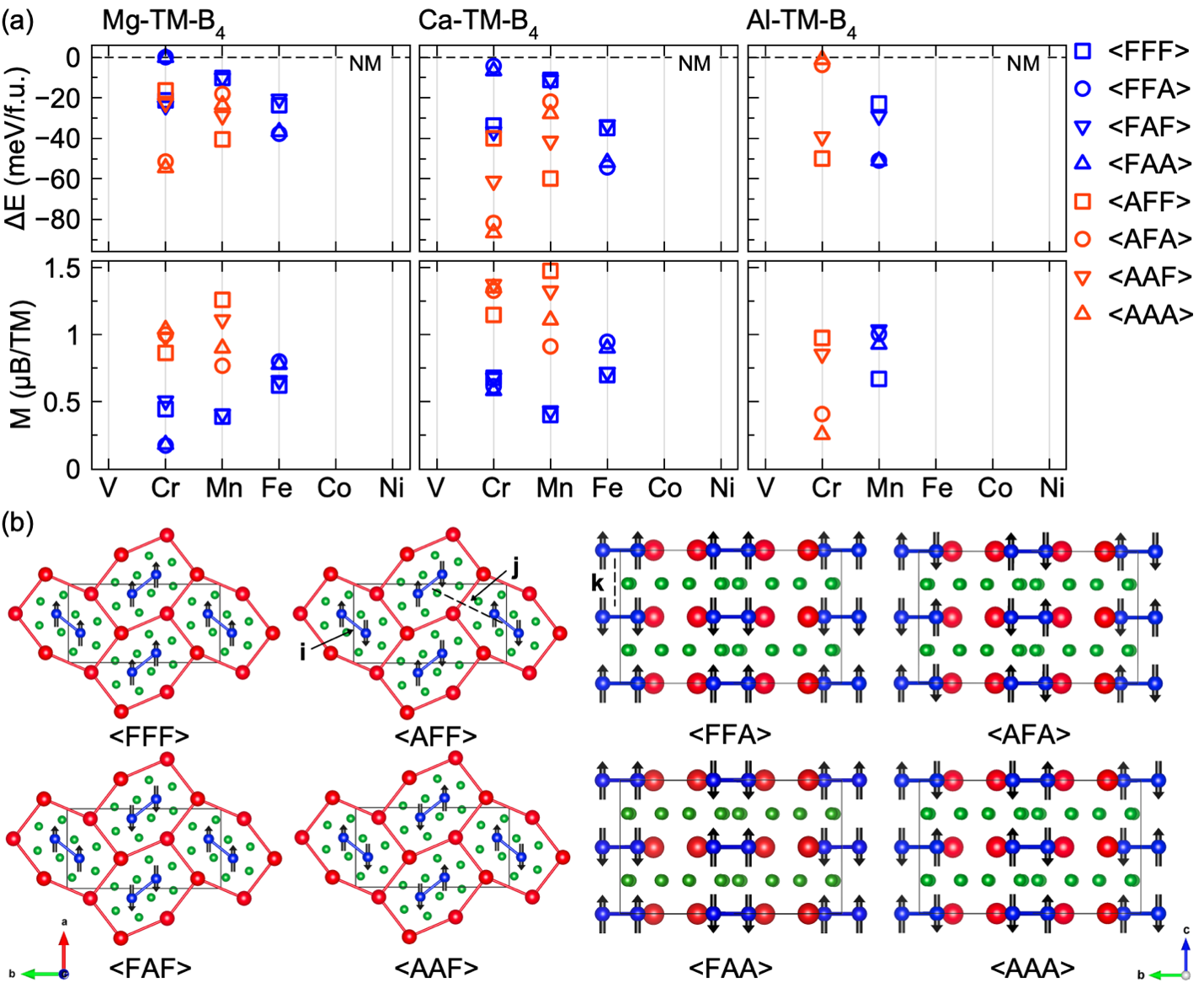}
	\caption{\textbf{The energy difference and magnetic moment for different magnetic states.} The symbols indicate different magnetic orders. The lower panel shows the magnetic order corresponding to different $<$\textbf{ijk}$>$ notations. Different colors show different elements: Mg (blue), Ca (red), and Al (green). The compounds with V, Co, Ni, and AlFeB$_4$ do not have magnetic solutions.}
	\label{fig5}
\end{figure}

In the 114 structure, one can consider three types of collinear magnetic order between T atoms. To clarify them, we use $<$\textbf{ijk}$>$ notation to represent the magnetic orders, where \textbf{i}, \textbf{j}, and \textbf{k} are either F (ferromagnetic) or A (antiferromagnetic). \textbf{i} represents the magnetic order within the dimer between two T atoms. \textbf{j} represents the magnetic order between two dimers in the $a$-$b$ plane, while \textbf{k} represents the magnetic order between layers of dimers along the $c$ direction. With this notation, there are eight different FM and AFM structures. Figure~\ref{fig5} shows all magnetic configurations for each compound with their magnetic energy and corresponding magnetic moments. In Mg- or Ca-based compounds, Cr, Fe, and Mn show magnetic states, while no magnetism is found in V, Co, and Ni systems. Different dopants result in different magnetic ground states. For instance, the ground states of MgCrB$_4$, MgMnB$_4$, and MgFeB$_4$ are $<$AAA$>$, $<$AFF$>$ and $<$FFA$>$, respectively. The energy of these states is about 40-80 meV/T lower than the NM state, which is like that of FM fcc Ni (60meV/Ni). The magnetic moments on transition metal atoms are typically close to 1 $\mu_{\rm{B}}$ in the ground states, larger than that in FM fcc Ni (0.6 $\mu_{\rm{B}}$). Thus, we expect these magnetic dimers' thermal stability to be very high as the Stoner temperature is expected to be like the one of Ni ($>$2500 K). Ca-based compounds show the same magnetic ground states as Mg-based compounds, suggesting very similar isoelectronic behavior (like the rigid band behavior of DOS discussed earlier). The magnetic behavior of Al systems appears to be similar to both Mg and Ca systems with the corresponding electronic population shifted by +1: AlCrB$_4$ has the same magnetic ground state as (Mg, Ca)MnB$_4$, and AlMnB$_4$ behaves as (Mg, Ca)FeB$_4$. The case of AlCrB$_4$ is less trivial due to the stable AFM insulating $<$AFF$>$ state. 

Next, we analyze the metastable magnetic states in Fig.~\ref{fig5}. In MgCrB$_4$ and CaCrB$_4$ all eight magnetic states can be stabilized. Nevertheless, Cr magnetism does not look localized. For all FM Cr dimers (\textbf{i}=F) the magnetic moments are much smaller, with the energies of  $<$FFA$>$ and $<$FAA$>$ states being very close to the NM state. Thus, individual Cr moment has a substantial degree of itineracy. However, the total magnetic moment of Cr dimer behaves like a very localized spin formation. Moreover, these AFM Cr dimers form AFM coupled ladders along the z-direction ($\textbf{i}$=A and $\textbf{k}$=A). The energy of the AFM ladder is well separated by $\sim$35 meV from the FM ladder ($\textbf{i}$=F and $\textbf{k}$=A). In Mg and Ca systems, such AFM Cr spin ladders weakly interact in plane as $<$AAA$>$ and $<$AFA$>$ spin configurations are nearly degenerate with very similar moments and energies. We can identify the magnetic ground state in these systems as weakly interacting AFM spin ladders, analog to spin glass systems. 

In MgMnB$_4$ , CaMnB$_4$ and isoelectronic AlCrB$_4$, corresponding Mn (Cr) atoms in dimer also form stable spin antiparallel order but form FM ladders as a ground state ($\textbf{i}$=A and $\textbf{k}$=F). While FM dimers have been found to be metastable their magnetic moments are much smaller, and the energies of such configurations appear close to the NM state. Thus, the atomic magnetic moments of Mn atoms in these compounds are also not localized and have rather significant itineracy, similar to Cr moments. However, Mn atoms in the dimer form very localized, AFM-coupled spin formation and the moments on Mn atoms in this formation practically do not depend on configuration relative to the nearest dimers. Overall, we expect that in Cr and Mn systems, the low-temperature spin disorder will be a disorder between weakly coupled spin ladders first, then at intermediate temperatures between more strongly interacting spin dimers inside each ladder; and at higher temperatures, the disorder should appear between spins inside dimer. This high-temperature disorder would also ultimately lead to the local moment disappearance on each atom. To some extent, such behavior would support the idea of spin clusters' disorder of Sokoloff  \cite{44}.

In MgFeB$_4$, CaFeB$_4$, and isoelectronic AlMnB$_4$, no AFM configurations between Fe atoms in the dimers are found (Fig.~\ref{fig5}a). It only allows the formation of FM dimers. In these systems, we predict the formation of stable AFM spin ladders along the z-direction. These ladders, however, weakly interact in plane so one can also expect that disordered spin ladders exist in plane at low temperatures.

\begin{figure}[b]
	\includegraphics[width=1\linewidth]{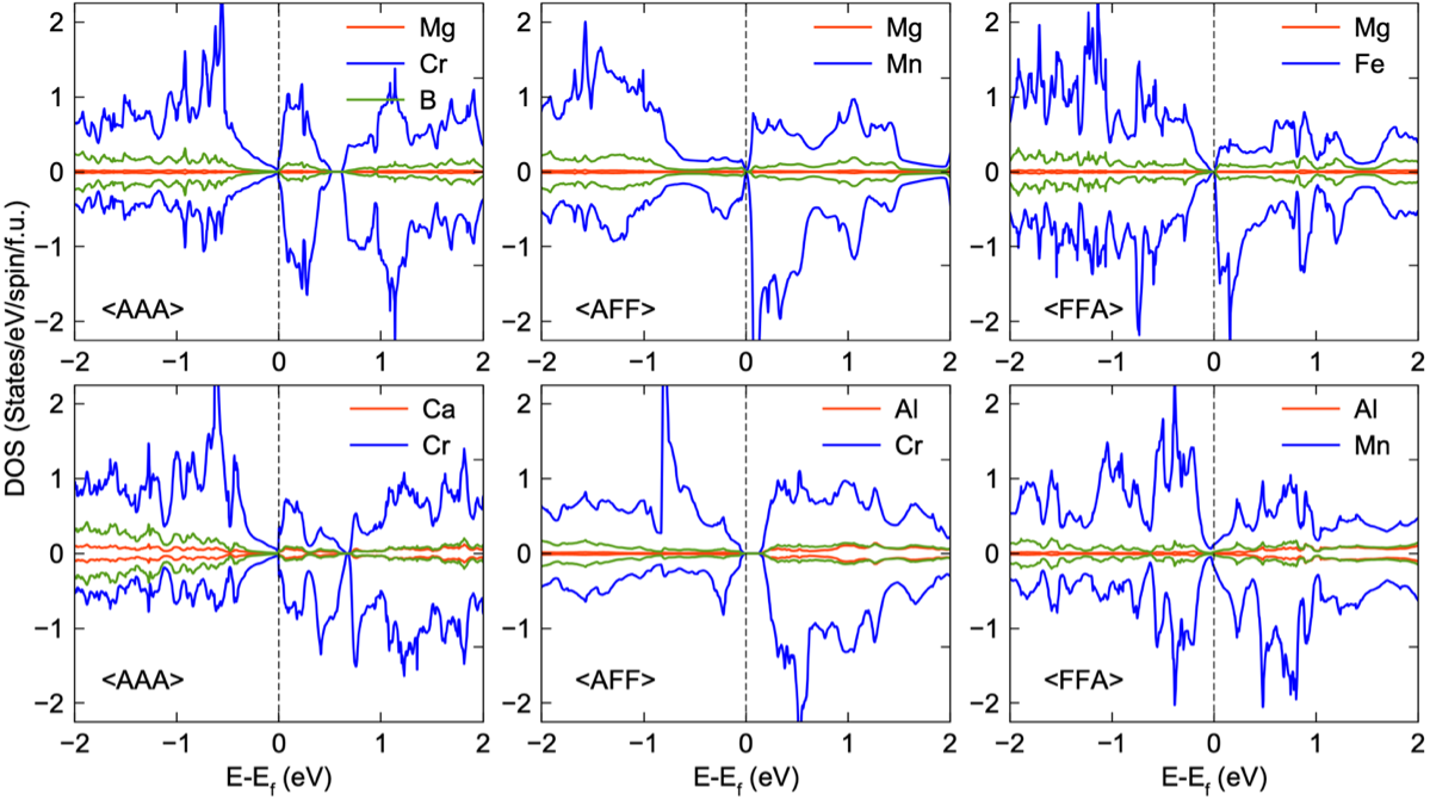}
	\caption{The spin-polarized density of states for magnetic ground states of $<$AAA$>$-MgCrB$_4$, $<$AFF$>$-MgMnB$_4$, $<$FFA$>$-MgFeB$_4$, $<$AAA$>$-CaCrB$_4$, $<$AFF$>$-AlCrB$_4$, and $<$FFA$>$-AlMnB$_4$.}
	\label{fig6}
\end{figure}

Figure~\ref{fig6} shows the spin-polarized DOS for magnetic ground states in Mg- and Al-based compounds. The DOS of Ca-based compounds are like Mg-based compounds, which can be seen by comparing the first column, i.e., $<$AAA$>$-MgCrB$_4$ and $<$AAA$>$-CaCrB$_4$. Compared to the NM DOS in Fig.~\ref{fig3}, the magnetism strongly reduces the $N(E_f)$, stabilizing the new ground states. The magnetic ground states all show very low $N(E_f)$, representing either semiconducting or weakly metallic states. In all cases, fluctuations are suppressed in the newly ordered ground magnetic state. The actual situation can be even more insulating as GGA/LDA methods traditionally tend to produce a metallic state that does not exist experimentally and underestimate the energy gap in semiconductors. Corresponding studies using methods like LDA+U or GW can be applied if the experimental data indicate larger band gaps in these materials.

Overall, while all nonmagnetic states appear to be metallic, magnetic interactions of density functional theory drive these metals to semiconductors or insulators. In this sense, the physics appears as the Slater metal-insulator transition proposed more than 70 years ago  \cite{45}. However, while the idea is theoretically very attractive, numerous experimental studies claimed that electron-electron interaction is more important for most materials than magnetic interactions. Thus, most known materials with metal-insulator transition follow strongly correlated scenarios or Mott behavior. Systems with Slater's metal-insulator transition are unique, and our prediction must be verified experimentally. The absence of magnetic states for Ni, Co, and V systems is most likely related to the fact that significant VHS near the Fermi level is not formed for the atoms of the beginning and the end of the $3d$ row.

\subsection{Superconductivity}
\begin{figure}[b]
	\includegraphics[width=1\linewidth]{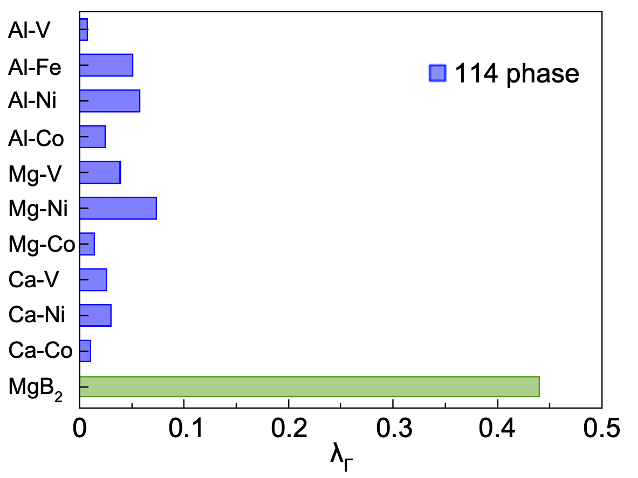}
	\caption{\textbf{Zone-center EPC strength for non-magnetic $\alpha$-ATB$_4$ phases.} Green is MgB$_2$ as a reference \cite{47}.}
	\label{fig7}
\end{figure}

We also examine the electron-phonon coupling (EPC) strength in the non-magnetic 114 systems, as metal borides can show phonon-mediated superconductivity  \cite{46}. We employ a recently developed frozen-phonon method to efficiently screen strong EPC candidates to compute the zone-center EPC strength \cite{47}. This method can identify the strong EPC candidates in MgB$_2$ and many metal borides because the zone-center EPC strongly correlates with these materials' full Brillouin zone EPC  \cite{47,48}. In Fig.~\ref{fig7}, we plot the zone-center EPC, $\lambda_\Gamma$, for the non-magnetic ground states of $\alpha$-ATB$_4$ compounds. We reference the zone-center EPC of MgB$_2$ computed in  \cite{47}. It shows that these 114 phases do not possess any strong EPC. Therefore, in these 114 systems, significant electron-phonon superconductivity should not be expected.

\begin{figure}[t]
	\includegraphics[width=1\linewidth]{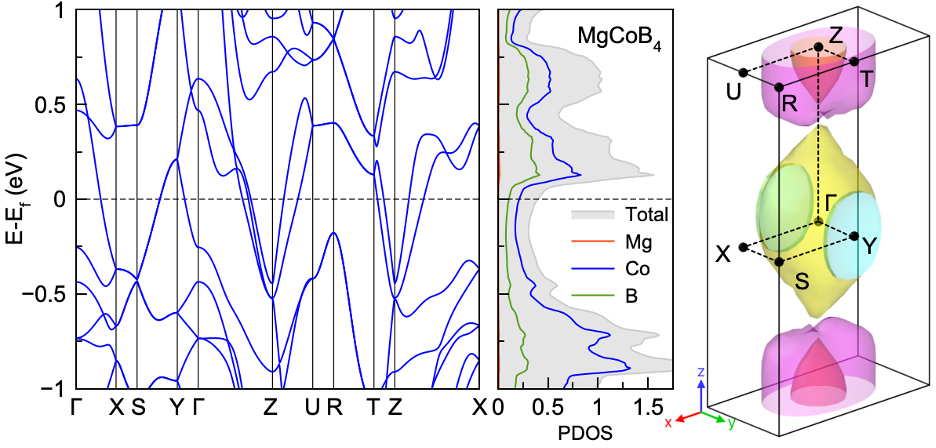}
	\caption{\textbf{Band structure, density of states, and Fermi surface of MgCoB$_4$.}}
	\label{fig8}
\end{figure}

However, we cannot dismiss the possibility of a different type of superconductivity pairing in these materials. While compounds with Cr and Mn can be eliminated from consideration of any superconductivity due to the presence of localized magnetic dimer states, the corresponding Fe (with Al) and all Co and Ni systems indeed represent metallic systems with the average value $N(E_f)$ close to the one in iron pnictides superconductors. Besides, the DOSs for all these systems (Fig.~\ref{fig3}) show certain electronic singularities near the Fermi levels, suggesting the possibility of different instabilities in the electronic, magnetic, or structural subsystems. Moreover, as we discussed above, it is evident that the closeness of the Fermi level to VHS (amplitude of electronic fluctuations) in these systems can be tuned by electronic (hole) doping. It is seen from the DOSs of MgCoB$_4$ and CaCoB$_4$ or AlFeB$_4$ and AlCoB$_4$ (Fig.~\ref{fig3}) where significant VHS exist right below and above the Fermi level. The Fermi surface of MgCoB$_4$ is shown in Fig.~\ref{fig8}. It forms anisotropic ellipsoids along the $c$ direction, i.e., the ladder direction of spin dimers. Such strong anisotropy in the Fermi surface can accompany superconductivity such as those observed in MgB$_2$ and cuprates  \cite{49,50}. Doping in the structure can move the VHS peak closer to the Fermi level, increasing $N(E_f)$ by 2-3 times. Thus, one can expect that the amplitude of electronic charge and spin fluctuations can be effectively manipulated in the needed way. Such tuning of the strength of spin fluctuations near the quantum critical point in these layered boride systems could represent a convenient playground for searching for spin fluctuation-mediated superconductivity  \cite{51}. 

\subsection{Synthesis exploration}
Due to exciting, predicted properties, we focused on the MgFeB$_4$ compound for synthetic exploration. Several factors make the synthesis of the MgFeB$_4$ phase challenging. Boron is a refractory element that requires a higher temperature to activate. Methods such as arc melting are commonly used to synthesize transition metal borides. In contrast, magnesium is a reactive element with high vapor pressure, $T_{\rm{boiling}}$ = 1100 $^{\circ}$C. This reactivity difference precludes the use of arc-melting or other high-temperature methods. We have recently shown the power of the mixing of refractory materials method when two refractory components are mixed by forming a binary compound via arc melting. The resulting compound is introduced into a reaction with more active components simultaneously and in close spatial proximity  \cite{52,53,54,55}. This method cannot be applied to Fe-B mixing due to the absence of boron-rich Fe borides. Melting of Fe+4B resulted in a mixture of FeB+3B, preventing homogeneous Fe-B mixing. The phase diagram shows that Mg and Fe metals are also immiscible in solid or liquid state  \cite{56}: Mg and Fe do not form binary compounds; the solubility of Mg in solid Fe below 1526 $^{\circ}$C is less than 0.6 at.\%; above 1526 $^{\circ}$C the segregation into two immiscible liquids of almost pure ($>$98 at.\%) Mg and Fe takes place. Our preliminary attempts to perform a reaction of elements in a wide temperature range of 600-1000 $^{\circ}$C were unsuccessful; they all produced a mixture of MgB$_2$, iron borides, and unreacted boron. 

Partial mixing of elements may be achieved when binary MgB$_2$ is used as a source of boron and magnesium. \textit{Gillan et al.} have shown that the reaction of MgB$_2$ and metal chlorides may produce corresponding binary metal borides, i.e., MgB$_2$+FeCl$_2$ = MgCl$_2$+FeB  \cite{57}. The formation of stable MgCl$_2$ was a thermodynamic driving force for this reaction. In our syntheses, the ternary Mg-Fe boride was a target, so we attempted a reaction of MgB$_2$+Fe. Yet, MgB$_2$ is a relatively inert precursor, and at temperatures of 750-850 $^{\circ}$C, the reaction of MgB$_2$ and elemental Fe is very slow, presumably due to high kinetic activation barriers.

\begin{figure}[t]
	\includegraphics[width=1\linewidth]{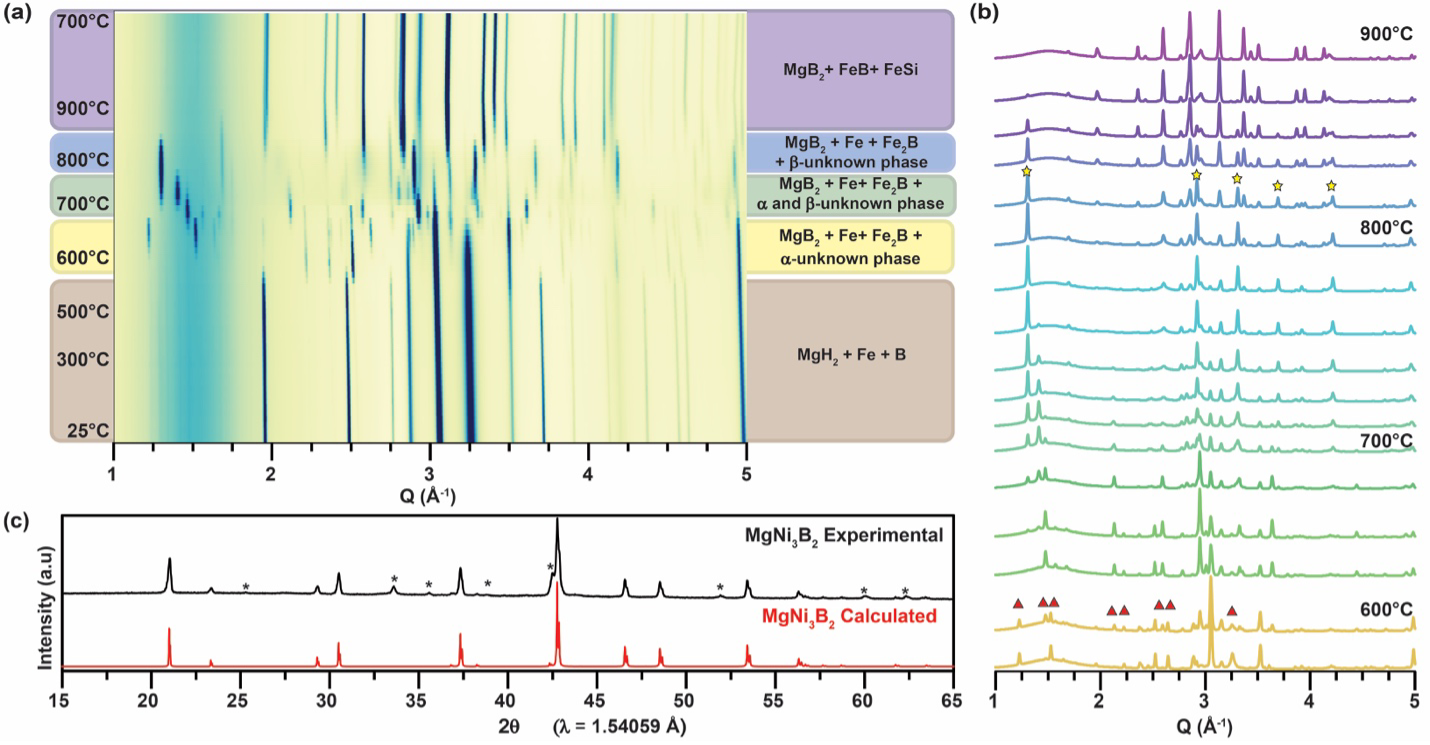}
	\caption{\textbf{Experimental attempts to synthesize 114 phases.} (a) \textit{In-situ} powder X-ray diffraction investigation of 1.3MgH$_2$ + Fe + 4B reaction in the 25-900$^{\circ}$C temperature range. Unindexed phases first appear at 560$^{\circ}$C for $\alpha$-phase and 690$^{\circ}$C for $\beta$-phase. (b) PXRD between the \textit{in-situ} reaction's temperature range of 560-900$^{\circ}$C. The red triangles represent unindexed peaks for the unknown phase $\alpha$ and the yellow stars represent unindexed peaks for the unknown phase $\beta$. (c) PXRD pattern, collected with Cu-$K_\alpha$ radiation, of ternary MgNi$_3$B$_2$ produced due to the reaction of MgB$_2$ and Ni in salt flux. Minor phase peaks of MgB$_2$ are represented by (*).}
	\label{fig9}
\end{figure}

Hence, we attempted a hydride reaction using MgH$_2$ as a source of Mg. Unlike ductile Mg, MgH$_2$ can be effectively mixed with Fe and B, and the released hydrogen may improve boron reactivity. The hydride approach's success was demonstrated by synthesizing several compounds in the Li-Ni-B system \cite{42,43,58,59}. To guide hydride syntheses, we performed an \textit{in-situ} powder X-ray diffraction (PXRD) study (Fig.~\ref{fig9}a). The formation of novel diffraction peaks was observed upon heating in 560-820 $^{\circ}$C range. However, they cannot be assigned to known or predicted ternary (including $\alpha$-MgFeB$_4$), binary phases, or elements in the Mg-Fe-B system. A set of intense unindexed peaks appeared at 560 $^{\circ}$C (assigned as an $\alpha$-unknown phase) and was present until 760 $^{\circ}$C (Fig.~\ref{fig9}b). Additionally, a second set of intense unknown peaks (assigned as $\beta$-unknown phase) formed at 690 $^{\circ}$C and disappeared at 820 $^{\circ}$C. Upon further heating, the formation of FeB and FeSi was observed. An \textit{in-situ} study was performed in a sealed silica capillary that can react with Mg at high temperatures as a Si or O source. The unknown phases were observed only in the \textit{in-situ} measurements at a narrow temperature range. Upon further heating, those phases decomposed such that the final products of the experiment contain no unknown phases, preventing elemental and spectroscopic analysis.

Nevertheless, from \textit{in-situ} data we can conclude that the unknown phases had limited thermal stability, and reactions at temperatures above 820 $^{\circ}$C are expected to produce binary FeB. Further identification of these unknown phases solely from powder diffraction data is challenging. Our \textit{ex-situ} hydride reactions do not yield new phases in any appreciable amounts. Some possibilities for this could be that the unknown phase incorporates silicon from the silica container, whereas for the \textit{ex-situ} reactions, niobium ampules are used for the reaction container. Another possibility is that this phase is metastable, and the slower heating ramp in \textit{ex-situ} reaction allows for this phase to decompose.

The thermodynamics calculations in Supplementary Note 1 suggest the MgFeB$_4$ is thermodynamically stable in an extensive temperature range. However, the synthesis is likely hindered by the formation of intermediate phases. These intermediates consume much reaction energy, leaving little driving force to form the target MgFeB$_4$. Such interplay between thermodynamics and kinetics represents a major challenge to control the solid-state synthesis  \cite{60,61}. These experimental results suggest that non-traditional solid-state methods may be required to synthesize MgFeB$_4$. The two main challenges are the immiscibility of Mg and Fe and the upper limit of $\sim$820 $^{\circ}$C synthetic temperature, which is relatively low for borides. When Fe was replaced with Ni, which is miscible with Mg, we were able to readily form a known ternary MgNi$_3$B$_2$ phase (Fig.~\ref{fig9}c). Further potential methods to produce target 114 borides discovered by the current computational study include flux reactions and designing of special precursors with pre-built functionality  \cite{62,63}. The latter approach was shown to be successful in the production of 1D B-P polymeric chains  \cite{64}. Flux reactions need to be carefully designed to dissolve all consisting elements (Mg, Fe, and B) yet not to react with container materials. Mg from flux readily reacts with silica, while salt fluxes react with Nb or Ta ampoule materials at elevated temperatures. Further synthetic explorations are currently underway.\\

\section{Conclusion}
In summary, using electronic structure calculations, we identified new structurally stable compounds among ternary borides of $\alpha$-ATB$_4$ type (A=Mg, Ca, Al; T=V, Cr, Mn, Fe, Ni, and Co). These predicted systems are characterized by numerous Van Hove singularities in their electronic spectrum formed by the emergence of highly localized quasimolecular states of $3d$ atomic dimers. Obtained VHS are stronger for electronic occupations near half filling of the $3d$ band and weaker for the states near the beginning and end of this band. The presence of such VHS, in turn, creates favorable conditions for the development of different types of fluctuations and the appearance of new quantum states. In our case, we analyzed magnetic instabilities. We found the formation of spin glass systems with spin dimers in magnetic semiconducting (or weakly metallic) states that the experiment can verify. The systematic appearance of VHS as a function of the electronic population should be verified by spectroscopic experiments as these localized states are well separated by 1-2 eV. Overall, we demonstrated how to search for the systems with developing Van Hove singularities. The proposed variation of the type of $3d$ atoms represents a convenient tuning of the VHS strength, which can lead to instability of the original paramagnetic Fermi liquid and the formation of new magnetic states. While producing MgFeB$_4$ experimentally was found to be challenging, we formulated a promising direction for further synthetic studies. The discovery of these materials would lead to an opportunity to study VHS systems and the possible formation of new quantum states, including unusual magnetic orders, superconducting states, charge density waves, and phase separation effects.\\

\section{Methods}
\noindent \textbf{First-principles calculations.} Density functional theory (DFT) calculations were carried out using the projector augmented wave (PAW) method \cite{65} implemented in the VASP code  \cite{66,67}. The exchange and correlation energy is treated with the generalized gradient approximation (GGA) and parameterized by the Perdew-Burke-Ernzerhof formula (PBE) \cite{68}. A plane-wave basis was used with a kinetic energy cutoff of 520 eV. The convergence criterion was set to 10$^{-5}$ eV for the total energy and 0.01 eV/$\rm{\AA}$ for ionic relaxation. Monkhorst-Pack's sampling scheme was adopted for Brillouin zone sampling with a k-point grid of 2$\pi$ $\times$ 0.033 $\rm{\AA}^{-1}$ for the structure optimization. Energy differences among different magnetic configurations and electronic density of states are computed with a denser k-point grid of 2$\pi$ $\times$ 0.022 $\rm{\AA}^{-1}$. Phonon calculations were performed with the density functional perturbation theory \cite{69} implemented in the VASP code and the Phonopy software \cite{70}. 	

\noindent \textbf{Phase stability calculations.} The phase stability is evaluated by the formation energy from spin-polarized calculations.  The formation energy $E_f$ of ATB$_4$ is calculated as $E_f=E-\frac{1}{6} E(\mathrm{A})-\frac{1}{6} E(\mathrm{T})-\frac{4}{6} E(\mathrm{B})$, where $E\left(\mathrm{M}_x \mathrm{N}_y \mathrm{B}_z\right)$ is the total energy of bulk ATB$_4$. $E(\mathrm{A})$, $E(\mathrm{T})$, and $E(\mathrm{B})$ are the total energies of A, T, and B ground-state bulk phases, respectively. $E_d$ is defined by the formation energy differences with respect to the three reference phases forming the Gibbs triangle on the convex hull (If $E_d$ = 0, it indicates the ATB$_4$ is a new stable phase, and the existing convex hull should be updated. The reference phases in the convex hulls are obtained from the Material Project database  \cite{40} and the OQMD database  \cite{71}. These reference phases are fully relaxed and the energies are re-calculated with the same DFT setting used in our high-throughput calculations.

\noindent \textbf{Electron-phonon coupling calculations.} The zone-center electron-phonon coupling ($\lambda_\Gamma$) was calculated using the difference between the screened ($\omega$) and unscreened ($\widetilde{\omega}$) zone-center phonon frequencies  \cite{47} as
\begin{equation}
\lambda_{\Gamma}=\frac{\widetilde{\omega}^2-\omega^2}{4 \omega^2}
\label{eq1}
\end{equation}
The screened phonon frequency was computed by fully self-consistent (SC) calculations in the displaced atomic configurations using the tetrahedron method with Bl\"{o}chl corrections. To compute the unscreened phonon frequency $\widetilde{\omega}$, the identical calculation was first performed in the equilibrium configuration, followed by the calculations with the displaced atoms, but with partial occupations fixed as the one in the equilibrium configuration. The detailed workflow of this method can be found in \cite{47}.

\noindent \textbf{Synthesis.} For experimental synthetic attempts, MgB$_2$ powder (Alfa Aesar, 99\%), Fe powder (JT Baker, 99.5\%), Cr powder (Alfa Aesar, 99.5\%), Mn powder (Alfa Aesar, 99.95\%), Ni powder (Alfa Aesar, 99.996\%), Co powder (Alfa Aesar, 99.998\%), MgH$_2$ powder (Alfa Aesar, 99\%), MgCl$_2$ powder (Alfa Aesar, 99\%), amorphous B powder (Alfa Asear, 98\%), and Nb ampoules were used. The total sample weight was 250 mg. A desired mixture of the ball-milled precursors was loaded into Nb ampoule, which was weld-shut under Ar atmosphere in the glovebox. The sealed Nb ampoules were enclosed in the Silica ampoule, which was evacuated and sealed using the hydrogen-oxygen torch. Various reactions were attempted: the reaction of three elements (Mg+Fe+4B), the reaction of magnesium boride with metal (2MgB$_2$+Fe), as well as hydride reactions (MgH$_2$+Fe+4B). The typical heating profile consists of a 10-hour ramp up to the desired temperature followed by isothermal annealing of 72 hours. After turning off the furnace, samples were allowed to cool back to room temperature. 

Powder X-ray Diffraction (PXRD). After annealing, all samples were exposed to air and ground into fine powder using agate mortar. PXRD characterization was performed using Rigaku MiniFlex600 powder diffractometer with Cu-$K_\alpha$ radiation and Ni-$K_\beta$ filter ($\lambda$ = 1.54059 $\rm{\AA}$). \textit{In-situ} PXRD was performed at beamline 17-BM at the Advanced Photon Source, Argonne National Laboratory ($\lambda$ = 0.24110 $\rm{\AA}$). A mixture of MgH$_2$, Fe, and B in a 1.3:1:4 ratio was ball-milled and loaded into a silica capillary with a 0.5 mm inner diameter and 0.7 mm outer diameter. The silica capillary was evacuated and flame-sealed such that the total length of the capillary was 50 mm. The capillary was placed vertically in a cell with resistive heating elements and aligned with the X-ray beam. Diffraction data were collected every 60 seconds as the sample was heated and cooled. Due to the reactor design, the thermocouple is slightly removed from the sample. We estimate that the error in the temperature is around 20-30 $^{\circ}$C. The reported temperature is the measured temperature. 

\section{Acknowledgments}
\noindent Y.S. acknowledges support from the Fundamental Research Funds for the Central Universities (20720230014). The work at Iowa State University was supported by National Science Foundation Awards No. DMR-2132666. Shaorong Fang from the Information and Network Center of Xiamen University is acknowledged for his help with high-performance computing.\\


\begin{thebibliography}{0}%
\makeatletter
\providecommand \@ifxundefined [1]{%
 \@ifx{#1\undefined}
}%
\providecommand \@ifnum [1]{%
 \ifnum #1\expandafter \@firstoftwo
 \else \expandafter \@secondoftwo
 \fi
}%
\providecommand \@ifx [1]{%
 \ifx #1\expandafter \@firstoftwo
 \else \expandafter \@secondoftwo
 \fi
}%
\providecommand \natexlab [1]{#1}%
\providecommand \enquote  [1]{``#1''}%
\providecommand \bibnamefont  [1]{#1}%
\providecommand \bibfnamefont [1]{#1}%
\providecommand \citenamefont [1]{#1}%
\providecommand \href@noop [0]{\@secondoftwo}%
\providecommand \href [0]{\begingroup \@sanitize@url \@href}%
\providecommand \@href[1]{\@@startlink{#1}\@@href}%
\providecommand \@@href[1]{\endgroup#1\@@endlink}%
\providecommand \@sanitize@url [0]{\catcode `\\12\catcode `\$12\catcode `\&12\catcode `\#12\catcode `\^12\catcode `\_12\catcode `\%12\relax}%
\providecommand \@@startlink[1]{}%
\providecommand \@@endlink[0]{}%
\providecommand \url  [0]{\begingroup\@sanitize@url \@url }%
\providecommand \@url [1]{\endgroup\@href {#1}{\urlprefix }}%
\providecommand \urlprefix  [0]{URL }%
\providecommand \Eprint [0]{\href }%
\providecommand \doibase [0]{http://dx.doi.org/}%
\providecommand \selectlanguage [0]{\@gobble}%
\providecommand \bibinfo  [0]{\@secondoftwo}%
\providecommand \bibfield  [0]{\@secondoftwo}%
\providecommand \translation [1]{[#1]}%
\providecommand \BibitemOpen [0]{}%
\providecommand \bibitemStop [0]{}%
\providecommand \bibitemNoStop [0]{.\EOS\space}%
\providecommand \EOS [0]{\spacefactor3000\relax}%
\providecommand \BibitemShut  [1]{\csname bibitem#1\endcsname}%
\let\auto@bib@innerbib\@empty
\end{thebibliography}%


\begin{thebibliography}{60}
\bibitem{1}	M. Fleck, A. M. Oleś, and L. Hedin, Magnetic Phases near the Van Hove Singularity in S- and d-Band Hubbard Models, Phys. Rev. B 56, 3159 (1997).
\bibitem{2}	J. González, Konh–Luttinger Superconductivity in Graphene, Phys. Rev. B 78, 205431 (2008).
\bibitem{3}	W. Kohn and J. M. Luttinger, New Mechanism for Superconductivity, Phys. Rev. Lett. 15, 524 (1965).
\bibitem{4}	T. M. Rice and G. K. Scott, New Mechanism for a Charge-Density-Wave Instability, Phys. Rev. Lett. 35, 120 (1975).
\bibitem{5}	L. Van Hove, The Occurrence of Singularities in the Elastic Frequency Distribution of a Crystal, Phys. Rev. 89, 1189 (1953).
\bibitem{6}	G. Li, A. Luican, J. M. B. Lopes dos Santos, A. H. Castro Neto, A. Reina, J. Kong, and E. Y. Andrei, Observation of Van Hove Singularities in Twisted Graphene Layers, Nat Phys 6, 109 (2010).
\bibitem{7}	S. Cho et al., Emergence of New van Hove Singularities in the Charge Density Wave State of a Topological Kagome Metal RbV3Sb5, Phys Rev Lett 127, 236401 (2021).
\bibitem{8}	S. Wu, Z. Zhang, K. Watanabe, T. Taniguchi, and E. Y. Andrei, Chern Insulators, van Hove Singularities and Topological Flat Bands in Magic-Angle Twisted Bilayer Graphene, Nat Mater 20, 488 (2021).
\bibitem{9}	Y. Hu, X. Wu, Y. Yang, S. Gao, N. C. Plumb, A. P. Schnyder, W. Xie, J. Ma, and M. Shi, Tunable Topological Dirac Surface States and van Hove Singularities in Kagome Metal GdV 6 Sn 6, Sci Adv 8, (2022).
\bibitem{10}	W. Wan, R. Harsh, P. Dreher, F. de Juan, and M. M. Ugeda, Superconducting Dome by Tuning through a van Hove Singularity in a Two-Dimensional Metal, NPJ 2D Mater Appl 7, 41 (2023).
\bibitem{11}	E. Dagotto, A. Nazarenko, A. Moreo, S. Haas, and M. Boninsegni, A Simple Theory for the Cuprates: The Antiferro-Magnetic van Hove Scenario, Journal of Superconductivity 8, 483 (1995).
\bibitem{12}	J. Bok and J. Bouvier, Van Hove Scenario for High T c Superconductors, in High Tc Superconductors and Related Transition Metal Oxides (Springer Berlin Heidelberg, Berlin, Heidelberg, 2007), pp. 35–41.
\bibitem{13}	V. Yu. Irkhin, A. A. Katanin, and M. I. Katsnelson, Effects of van Hove Singularities on Magnetism and Superconductivity in the t-t' Hubbard Model: A Parquet Approach, Phys Rev B 64, 165107 (2001).
\bibitem{14}	W. F. Goh and W. E. Pickett, A Mechanism for Weak Itinerant Antiferromagnetism: Mirrored van Hove Singularities, EPL (Europhysics Letters) 116, 27004 (2016).
\bibitem{15}	V. P. Antropov, M. I. Katsnelson, V. G. Koreshkov, A. I. Likhtenstein, A. V. Trefilov, and V. G. Vaks, On Anomalies of Anisotropic Thermal Expansion near Points of Electronic Topological Transitions in Noncubic Metals. Applications to Zn, Cd and Cd-Mg Alloys, Phys Lett A 130, 155 (1988).
\bibitem{16}	V. I. Antropov, V. G. Vaks, M. I. Katsnel’son, V. G. Koreshkov, A. I. Likhtenshteĭn, and A. V Trefilov, Effect of Proximity of the Fermi Level to Singular Points in the Band on the Kinetic and Lattice Properties of Metals and Alloys, Soviet Physics Uspekhi 31, 278 (1988).
\bibitem{17}	C. W. Hicks et al., Strong Increase of T c of Sr 2 RuO 4 Under Both Tensile and Compressive Strain, Science (1979) 344, 283 (2014).
\bibitem{18}	A. Steppke et al., Strong Peak in T c of Sr 2 RuO 4 under Uniaxial Pressure, Science (1979) 355, (2017).
\bibitem{19}	M. E. Barber, A. S. Gibbs, Y. Maeno, A. P. Mackenzie, and C. W. Hicks, Resistivity in the Vicinity of a van Hove Singularity: <math Display="inline"> <mrow> <msub> <mrow> <mi>Sr</Mi> </Mrow> <mrow> <mn>2</Mn> </Mrow> </Msub> <msub> <mrow> <mi>RuO</Mi> </Mrow> <mrow> <mn>4</Mn> </Mrow> </Msub> </Mrow> </Math> under Uniaxial Pressure, Phys Rev Lett 120, 076602 (2018).
\bibitem{20}	L. Liu, C. Wang, L. Zhang, C. Liu, C. Niu, Z. Zeng, D. Ma, and Y. Jia, Surface Van Hove Singularity Enabled Efficient Catalysis in Low-Dimensional Systems: CO Oxidation and Hydrogen Evolution Reactions, Journal of Physical Chemistry Letters 13, 740 (2022).
\bibitem{21}	Yu. B. Kuzma, Crystal Structure of the Compound YCrB4 and Its Analogs, Sov. Phys. Crystallogr. 15, 312 (1970).
\bibitem{22}	Yu. B. Kuzma, New Ternary Compounds with the Structure of YCrB4 Type, Dopov. Akad. Nauk Ukr. RSR Ser. A 8, 756 (1970).
\bibitem{23}	P. Rogl, New Ternary Borides with YCrB4-Type Structure, Mater Res Bull 13, 519 (1978).
\bibitem{24}	R. Sobczak and P. Rogl, Magnetic Behavior of New Ternary Metal Borides with YCrB4-Type Structure, J Solid State Chem 27, 343 (1979).
\bibitem{25}	I. Veremchuk, T. Mori, Yu. Prots, W. Schnelle, A. Leithe-Jasper, M. Kohout, and Yu. Grin, Synthesis, Chemical Bonding and Physical Properties of RERhB4 (RE=Y, Dy–Lu), J Solid State Chem 181, 1983 (2008).
\bibitem{26}	T. Mori, S. Okada, and K. Kudou, Magnetic Properties of Thulium Aluminoboride TmAlB4, J Appl Phys 97, 10A910 (2005).
\bibitem{27}	S. Okada, T. Shishido, T. Mori, K. Kudou, K. Iizumi, T. Lundström, and K. Nakajima, Properties of REAlB4 and Lu2AlB6 Crystals Grown from RE–Al–B (RE=Tm, Yb, Lu) Melts, J Alloys Compd 408–412, 547 (2006).
\bibitem{28}	R. T. Macaluso, S. Nakatsuji, K. Kuga, E. L. Thomas, Y. Machida, Y. Maeno, Z. Fisk, and J. Y. Chan, Crystal Structure and Physical Properties of Polymorphs of LnAlB 4 (Ln ) Yb, Lu), Chem. Mater. 19, 1918 (2007).
\bibitem{29}	T. Mori, H. Borrmann, S. Okada, K. Kudou, A. Leithe-Jasper, U. Burkhardt, and Yu. Grin, Crystal Structure, Chemical Bonding, Electrical Transport, and Magnetic Behavior of TmAlB4, Phys Rev B 76, 064404 (2007).
\bibitem{30}	S. Nakatsuji et al., Superconductivity and Quantum Criticality in the Heavy-Fermion System $\beta$-YbAlB4, Nat Phys 4, 603 (2008).
\bibitem{31}	K. Kuga, Y. Karaki, Y. Matsumoto, Y. Machida, and S. Nakatsuji, Superconducting Properties of the Non-Fermi-Liquid System $\beta$-YbAlB4, Phys Rev Lett 101, 137004 (2008).
\bibitem{32}	F. Dai, Z. Feng, and Y. Zhou, First Principles Investigation on Mechanical and Thermal Properties of $\alpha$‐ and $\beta$‐YAlB4 Ultra‐high Temperature Ceramics, Journal of the American Ceramic Society 101, 5694 (2018).
\bibitem{33}	H. Orsini-Rosenberg and W. Steurer, Ab Initio Investigations on the Stability of Seven-Fold Approximants, Philosophical Magazine 91, 2567 (2011).
\bibitem{34}	A. Candan, G. Surucu, and A. Gencer, Electronic, Mechanical and Lattice Dynamical Properties of YXB4 (X = Cr, Mn, Fe, and Co) Compounds, Phys Scr 94, 125710 (2019).
\bibitem{35}	Y. J. Colón and R. Q. Snurr, High-Throughput Computational Screening of Metal–Organic Frameworks, Chem Soc Rev 43, 5735 (2014).
\bibitem{36}	A. A. Emery and C. Wolverton, High-Throughput DFT Calculations of Formation Energy, Stability and Oxygen Vacancy Formation Energy of ABO3 Perovskites, Sci Data 4, 170153 (2017).
\bibitem{37}	W. Sun et al., A Map of the Inorganic Ternary Metal Nitrides, Nature Materials 2019 18:7 18, 732 (2019).
\bibitem{38}	J. Noh, G. H. Gu, S. Kim, and Y. Jung, Machine-Enabled Inverse Design of Inorganic Solid Materials: Promises and Challenges, Chem Sci 11, 4871 (2020).
\bibitem{39}	K. Yubuta, T. Mori, A. Leithe-Jasper, Y. Grin, S. Okada, and T. Shishido, Direct Observation of the Intergrown $\alpha$-Phase in $\beta$-TmAlB4 via High-Resolution Electron Microscopy, Mater Res Bull 44, 1743 (2009).
\bibitem{40}	A. Jain et al., Commentary: The Materials Project: A Materials Genome Approach to Accelerating Materials Innovation, APL Mater 1, 011002 (2013).
\bibitem{41}	W. Sun, S. T. Dacek, S. P. Ong, G. Hautier, A. Jain, W. D. Richards, A. C. Gamst, K. A. Persson, and G. Ceder, The Thermodynamic Scale of Inorganic Crystalline Metastability, Sci Adv 2, e1600225 (2016).
\bibitem{42}	V. Gvozdetskyi et al., Computationally Driven Discovery of a Family of Layered LiNiB Polymorphs, Angewandte Chemie - International Edition 58, 15855 (2019).
\bibitem{43}	G. Bhaskar et al., Topochemical Deintercalation of Li from Layered LiNiB: Toward 2D MBene, J Am Chem Soc 143, 4213 (2021).
\bibitem{44}	J. B. Sokoloff, A Droplet Model for Ferromagnetic Spin Waves above Tc, Phys Rev B 17, 2380 (1978).
\bibitem{45}	J. C. Slater, Magnetic Effects and the Hartree-Fock Equation, Physical Review 82, 538 (1951).
\bibitem{46}	J. Nagamatsu, N. Nakagawa, T. Muranaka, Y. Zenitani, and J. Akimitsu, Superconductivity at 39 K in Magnesium Diboride, Nature 410, 63 (2001).
\bibitem{47}	Y. Sun, F. Zhang, C.-Z. Wang, K.-M. Ho, I. I. Mazin, and V. Antropov, Electron-Phonon Coupling Strength from Ab Initio Frozen-Phonon Approach, Phys Rev Mater 6, 074801 (2022).
\bibitem{48}	R. Wang, Y. Sun, F. Zhang, F. Zheng, Y. Fang, S. Wu, H. Dong, C.-Z. Wang, V. Antropov, and K.-M. Ho, High-Throughput Screening of Strong Electron–Phonon Couplings in Ternary Metal Diborides, Inorg Chem 61, 18154 (2022).
\bibitem{49}	A. Y. Liu, I. I. Mazin, and J. Kortus, Beyond Eliashberg Superconductivity in MgB2: Anharmonicity, Two-Phonon Scattering, and Multiple Gaps, Phys Rev Lett 87, 087005 (2001).
\bibitem{50}	T. P. Devereaux, T. Cuk, Z. X. Shen, and N. Nagaosa, Anisotropic Electron-Phonon Interaction in the Cuprates, Phys Rev Lett 93, 117004 (2004).
\bibitem{51}	R. Wang, Y. Sun, V. Antropov, Z. Lin, C. Z. Wang, and K. M. Ho, Theoretical Prediction of a Highly Responsive Material: Spin Fluctuations and Superconductivity in FeNiB2 System, Appl Phys Lett 115, 182601 (2019).
\bibitem{52}	G. Akopov et al., Synthesis-Enabled Exploration of Chiral and Polar Multivalent Quaternary Sulfides, Chem Sci 12, 14718 (2021).
\bibitem{53}	S. J. Lee, J. Won, L. L. Wang, D. Jing, C. P. Harmer, J. Mark, G. Akopov, and K. Kovnir, New Noncentrosymmetric Tetrel Pnictides Composed of Square-Planar Gold(I) with Peculiar Bonding, Chemistry – A European Journal 27, 7383 (2021).
\bibitem{54}	G. Akopov et al., Third Time’s the Charm: Intricate Non-Centrosymmetric Polymorphism in LnSiP3 (Ln = La and Ce) Induced by Distortions of Phosphorus Square Layers, Dalton Transactions 50, 6463 (2021).
\bibitem{55}	S. J. Lee, G. Viswanathan, S. L. Carnahan, C. P. Harmer, G. Akopov, A. J. Rossini, G. J. Miller, and K. Kovnir, Add a Pinch of Tetrel: The Transformation of a Centrosymmetric Metal into a Nonsymmorphic and Chiral Semiconductor, Chemistry – A European Journal 28, e202104319 (2022).
\bibitem{56}	A. A. Nayeb-Hashemi and J. B. Clark, Phase Diagrams of Binary Magnesium Alloys (ASM International, 1988).
\bibitem{57}	J. P. Abeysinghe, A. F. Kölln, and E. G. Gillan, Rapid and Energetic Solid-State Metathesis Reactions for Iron, Cobalt, and Nickel Boride Formation and Their Investigation as Bifunctional Water Splitting Electrocatalysts, ACS Materials Au 2, 489 (2022).
\bibitem{58}	G. Bhaskar et al., Path Less Traveled: A Contemporary Twist on Synthesis and Traditional Structure Solution of Metastable LiNi12B8, ACS Materials Au 2, 614 (2022).
\bibitem{59}	V. Gvozdetskyi, M. P. Hanrahan, R. A. Ribeiro, T. H. Kim, L. Zhou, A. J. Rossini, P. C. Canfield, and J. V. Zaikina, A Hydride Route to Ternary Alkali Metal Borides: A Case Study of Lithium Nickel Borides, Chemistry – A European Journal 25, 4123 (2019).
\bibitem{60}	N. J. Szymanski et al., Understanding the Fluorination of Disordered Rocksalt Cathodes through Rational Exploration of Synthesis Pathways, Chemistry of Materials 34, 7015 (2022).
\bibitem{61}	M. Bianchini et al., The Interplay between Thermodynamics and Kinetics in the Solid-State Synthesis of Layered Oxides, Nature Materials 2020 19:10 19, 1088 (2020).
\bibitem{62}	M. G. Kanatzidis, R. Pöttgen, and W. Jeitschko, The Metal Flux: A Preparative Tool for the Exploration of Intermetallic Compounds, Angewandte Chemie International Edition 44, 6996 (2005).
\bibitem{63}	P. C. Canfield, New Materials Physics, Reports on Progress in Physics 83, 016501 (2019).
\bibitem{64}	K. E. Woo, J. Wang, J. Mark, and K. Kovnir, Directing Boron-Phosphorus Bonds in Crystalline Solid: Oxidative Polymerization of P=B=P Monomers into 1D Chains, J Am Chem Soc 141, 13017 (2019).
\bibitem{65}	P. E. Blöchl, Projector Augmented-Wave Method, Phys Rev B 50, 17953 (1994).
\bibitem{66}	G. Kresse and J. Furthmüller, Efficiency of Ab-Initio Total Energy Calculations for Metals and Semiconductors Using a Plane-Wave Basis Set, Comput Mater Sci 6, 15 (1996).
\bibitem{67}	G. Kresse and J. Furthmüller, Efficient Iterative Schemes for Ab Initio Total-Energy Calculations Using a Plane-Wave Basis Set, Phys Rev B 54, 11169 (1996).
\bibitem{68}	J. P. Perdew, K. Burke, and M. Ernzerhof, Generalized Gradient Approximation Made Simple, Phys Rev Lett 77, 3865 (1996).
\bibitem{69}	S. Baroni, S. de Gironcoli, A. Dal Corso, and P. Giannozzi, Phonons and Related Crystal Properties from Density-Functional Perturbation Theory, Rev Mod Phys 73, 515 (2001).
\bibitem{70}	A. Togo and I. Tanaka, First Principles Phonon Calculations in Materials Science, Scr Mater 108, 1 (2015).
\bibitem{71}	S. Kirklin, J. E. Saal, B. Meredig, A. Thompson, J. W. Doak, M. Aykol, S. Rühl, and C. Wolverton, The Open Quantum Materials Database (OQMD): Assessing the Accuracy of DFT Formation Energies, NPJ Comput Mater 1, 15010 (2015).


\end{thebibliography}
\bibliographystyle{apsrev4-1}

\end{document}